# Multiple testing procedures under confounding

Debashis Ghosh[*,1]

*Pennsylvania State University*

**Abstract:** While multiple testing procedures have been the focus of much statistical research, an important facet of the problem is how to deal with possible confounding. Procedures have been developed by authors in genetics and statistics. In this chapter, we relate these proposals. We propose two new multiple testing approaches within this framework. The first combines sensitivity analysis methods with false discovery rate estimation procedures. The second involves construction of shrinkage estimators that utilize the mixture model for multiple testing. The procedures are illustrated with applications to a gene expression profiling experiment in prostate cancer.

## 1. Introduction

In the study of complex diseases such as cancer, investigators have sought to implicate candidate gene polymorphisms with disease. Candidate gene polymorphisms have been typically identified in case-control studies, where their frequencies in cases and controls are typically compared. Such studies are quite commonplace in the scientific literature.

For this context, we focus on the problem of multiple testing. A useful quantity to consider in multiple testing situations is the false discovery rate (FDR) (Benjamini and Hochberg [2]). Sabatti et al. [24] and Abecasis et al. [1] have advocated its use in genetic association studies. There has been a lot of work on statistical procedures involving the false discovery rate, which we review in Section 2. While much work has been done on multiple testing procedures, very little work to date has been done on estimation for this setting.

Recently, Efron [11] has argued that these methods presume theoretical null distributions, which might be incorrect, and has instead argued for use of an empirical null distribution. A parallel development of this problem has occurred in genetic association studies (Cardon and Bell [7]), in which statistical geneticists look at frequencies of alleles in affected (case) and unaffected (control) populations to determine associations between genes and disease. Wright [36] argued that a variety of population genetic forces, such as nonrandom mating and genetic drift, might induce genetically similar subgroups within populations. This is referred to as population structure in the genetic literature. Proposals for addressing this problem have been developed by Devlin and Roeder [10] and Rosenberg and Pritchard [23].

---

*Supported by the National Institutes of Health and the National Science Foundation.

[1]Department of Statistics, The Pennsylvania State University, 326 Thomas Building, Ann Arbor, MI 48109-2029, USA, e-mail: debashisorama@gmail.com

In this article, we seek to unify these proposals in a framework for multiple testing in which we adjust for confounding. The population structure that must be accounted for in genetic association studies represents one type of confounding. There has also been mention on the role of other confounders in differential expression analyses for microarray data (Bhattacharya et al. [6] and Ghosh and Chinnaiyan [17]). Confounding is a problem in that it biases the associations that are observed in the data. In other words, because of confounding, under the null hypothesis of no association, the usual test statistic does not have the correct mean.

The structure of this paper is as follows. In Section 2, we define the false discovery rate and review the standard mixture model for multiple hypothesis testing that has been used in the literature. We then slightly generalize the model to allow for confounding and then relate three proposals for multiple testing (Devlin and Roeder [10], Pritchard and Rosenberg [23] and Efron [11]) to this model. We then propose two analytical schemes for multiple testing in the presence of confounding. The first involves utilization of sensitivity analysis methods to adjust for confounding (Lin et al. [19]), followed by q-value estimation (Storey [28]) or a false discovery rate controlling procedure. This, along with the link to related shrinkage estimation procedures (James and Stein [18], Sen and Saleh [26, 27] and George [15]), is discussed in Section 3. In Section 4, we discuss estimation of association, along with confidence intervals in the multiple testing situation in the presence of confounding. We highlight the procedures with applications to a gene expression data from a prostate cancer study in Section 5. Finally, we conclude with some discussion in Section 6.

## 2. Data structures, background and preliminaries

Suppose there are $g$ genes that we wish to make gene-specific hypotheses about. We can cross-classify the corresponding $g$ hypotheses into their true status (i.e. null is true or alternative is true) and also based on whether or not we decide to reject the null hypothesis. Such a table is given in Table 1.

The classical quantity that has been controlled in multiple testing problems has been the familywise type I error rate (FWER). Based on Table 1, the FWER is defined as $P(V \geq 1)$. By contrast, the definition of false discovery rate (FDR) as put forward by Benjamini and Hochberg [2] is

$$FDR \equiv E\left[\frac{V}{Q} \mid Q > 0\right] P(Q > 0).$$

The conditioning on the event $[Q > 0]$ is needed because the fraction $V/Q$ is not well-defined when $Q = 0$. If all the null hypotheses are true (i.e., $g_0 = g$), then control of the false discovery also provides control of the familywise type I error rate. In general, however, control of the two quantities are not equivalent, and the control of the false discovery rate is less conservative than that of FWER.

Table 1
*Outcomes of G tests of hypotheses*

|  | Accept | Reject | Total |
|---|---|---|---|
| True null | U | V | $G_0$ |
| True alternative | T | S | $G_1$ |
|  | W | Q | $G$ |

FDR-related procedures fall into two classes: (1) methods that aim to control FDR; (2) methods for direct estimation of FDR based on a fixed rejection region. The first class of procedures has been proposed by many authors, such as Benjamini and Hochberg [2], Benjamini and Liu [3], Benjamini and Yekutieli [4] and Sarkar [25]. Efron et al. [12], Storey [28, 29] and Genovese and Wasserman [13], 2005 have developed the second class of procedures. The two classes of methods have been unified by Storey et al. [30] and Genovese and Wasserman (2005). The FDR is also linked intimately to the q-value (Storey [28]), the smallest FDR at which a null hypothesis can be rejected.

## 2.1. Mixture model framework for multiple testing

For false discovery rate estimation procedures, many authors have studied a mixture model for multiple testing. To test each null hypothesis, let $T_g$ denote the test statistic for the $g$th gene. Define indicator variables $H_1, \ldots, H_G$ corresponding to $T_1, \ldots, T_G$, where $H_i = 0$ if the null hypothesis is true and $H_i = 1$ if the alternative hypothesis is true. Assume that $H_1, \ldots, H_G$ are a random sample from a Bernoulli distribution where $P(H_i = 0) = \pi_0$, $i = 1, \ldots, G$. We define the densities $f_0$ and $f_1$ corresponding to $T_i|H_i = 0$ and $T_i|H_i = 1$, $(i = 1, \ldots, m)$. The marginal density of the test statistics $T_1, \ldots, T_m$ is given by

$$(2.1) \qquad f(t) \equiv \pi_0 f_0(t) + (1 - \pi_0) f_1(t).$$

The mixture model framework represented in (2.1) has been used by several authors to study the false discovery rate (e.g., Efron et al. [12], Storey [28] and Genovese and Wasserman [14]). Given an estimate of $\pi_0$, methods for false discovery rate estimation have been developed by several authors (Efron et al. [12] and Storey [28]). There are several methods for estimating $\pi_0$ in the literature, most of which are based on the defintion of $\pi_0$ as the derivative of the density of test statistics corresponding to the null hypothesis evaluated at one. The latter definition can be found in Storey and Tibshirani [31]; methods of estimation of $\pi_0$ can be found in Storey [28], Pounds and Cheng [22] and Dalmasso et al. [8]. Intuitively, it is estimated based on p-values that in a region close to one; the size of the region will be inevitably linked to a bias-variance tradeoff regarding the estimate of $\pi_0$.

## 2.2. Adjustments for confounding

In model (2.1), authors have usually assumed $f_0$ to be known. For example, if the test statistics are p-values, then $f_0$ has been assumed to be the pdf of a Uniform(0,1) random variable. If the statistics are Wald-type statistics (e.g., two sample t-tests), then $f_0$ is usually taken to be a normal distribution with mean zero and variance one.

There have been certain proposals in which authors attempted to use permutation methods to estimate $f_0$ (e.g., Efron et al. [12]). This technique has been used to account for correlation between the hypotheses, e.g. genes on a microarray having correlation. Permutation is often carried out by interchanging labels between independent samples. However, a crucial assumption of permutation method is that under the null hypothesis, all assignments of class labels to samples are exchangeable. In the presence of confounding, this assumption is no longer true. This is why Efron [11] writes that "permutation methods do no automatically resolve the issue of the theoretical versus the empirical null hypothesis."

To account for confounding, let us assume that under the null hypothesis, the $i$th test statistic $T_i$ has distribution $f_{0i}^C(t)$, where the subscript $C$ denotes confounding. Assume that under the alternative hypothesis, the distribution is $f_{1i}^C(t)$ for $T_i$, $i = 1, \ldots, n$. Then we have the following mixture model:

$$T_i \stackrel{ind}{\sim} \pi_0 f_{0i}^C(t) + (1 - \pi_0) f_{1i}^C(t). \tag{2.2}$$

Assume further that $f_{1i}^C(t)$ does not depend on $i$. Suppose there were supplementary test statistics $U_1, \ldots, U_K$ available for which it were known that the null hypothesis was true. The framework of Devlin and Roeder [10] works in the following manner. Using a population genetic model and the presence of population stratification, they show that $f_{0i}^C(t)$ will be relatively overdispersed compared to $f_{0i}(t)$, which they assume not to depend on $i$. To be specific, $f_{0i}^C(t) = \sigma_i^{-1} f_0(t/\sigma_i)$. While there should also be a mean shift (i.e. bias correction) in $f_0^C$ relative to $f_0$, Devlin and Roeder [10] argue that the overdispersion outweighs the bias; this has been supported by other authors using simulation studies (Wacholder et al. [34]). Making a further assumption that the scale parameter $\sigma_i$ is constant across the genome so that it no longer depends on $i$, Devlin and Roeder [10] use the statistics $U_1, \ldots, U_K$ to estimate $\sigma$. Let the resulting estimator be denoted as $\hat{\sigma}$; one can then use in $\hat{\sigma}^{-1} f_0(t/\hat{\sigma})$ as the reference null hypothesis for multiple testing. Devlin and Roeder [10] argue for the constant variance assumption across the genome using population genetic arguments.

An alternative approach to adjusting for confounding was put forward by Pritchard and Rosenberg [23]. They postulated the existence of a latent variable $C$ such that conditional on $C = c$,

$$T_i | C = c \stackrel{iid}{\sim} \pi_0 f_0(t) + (1 - \pi_0) f_1(t), \tag{2.3}$$

i.e. within subpopulations defined by $C$, the test statistics have the distribution defined as in (2.1). In their method, what is needed are supplementary data that can be used to define $C$ for each individual. Just as in the genomic control method of Devlin and Roeder [10], these are data on genes for which the null hypothesis is known to be true. The approach of Pritchard and Rosenberg [23] is to infer subpopulation membership ($C$) using the supplementary data and then do standard multiple testing within each subpopulation. Conditional on $C$ (i.e., within subpopulations defined by $C$), the theoretical null and alternative distributions can be used for the multiple testing problem. For Pritchard and Rosenberg [23], they use the supplemental data to infer subpopulation membership and then for those defined by common values for the inferred cluster membership, they can perform the standard multiple testing procedures.

The approach of Efron [11] involves the following model:

$$T_i \stackrel{iid}{\sim} \pi_0 f_0^C(t) + (1 - \pi_0) f_1^C(t). \tag{2.4}$$

Thus, the density of the test statistics corresponding to the null and alternative hypotheses are left unspecified. No relationship between $f_0^C(t)$ to $f_0(t)$ is formulated, making this approach more flexible than that of Devlin and Roeder [10]. In addition, no supplemental data are required on known true null hypotheses, in contrast to the Devlin and Roeder [10] and Pritchard and Rosenberg [23] proposals. To make estimation in (2.4) feasible, Efron [11] makes a zero-matching assumption, namely that all statistics near zero come from the null hypothesis component and relates $f_1$ to $f_0$ via an exponential tilt (Lindsey [20]).

In comparing the proposals, we find that both the method of Pritchard and Rosenberg [23] and Devlin and Roeder [10] both require supplemental data, as well as an assumption that the supplemental data are not associated with the phenotype in any manner. By contrast, no supplemental data are needed for the method of Efron [11]. However, a desirable feature of the first two proposals mentioned is that they explicitly model the confounding in an easily interpretable way into the model. The Efron approach is more algorithmic in nature, although he mentions confounding as being one reason to prefer the empirical null relative to the theoretical null hypothesis.

If we think of confounding as manifesting through a bias and variance adjustment to the theoretical null distribution, then we find that the genomic control method makes a variance adjustment to the theoretical null, while the approach of Pritchard and Rosenberg [23] proposal makes an adjustment to both through the latent variable $C$. The Efron [11] procedure also does the same without specifying a probabilistic model for the confounding. The first approach we propose in this paper is to postulate confounding in a manner analogous to Pritchard and Rosenberg [23]; however, the actual multiple testing procedures do not require additional data, which makes it similar to the approach of Efron [11]. An important step in application of the proposed methodology is the use of sensitivity analysis methods, which are described next.

## 3. Sensitivity analysis and multiple testing methodologies

To incorporate confounding into the analysis, we will utilize the framework of Lin et al. [19]. They cast sensitivity analysis into a regression modelling framework in which the confounder was treated as an unobserved covariate. While they focused primarily on binary and time-to-event outcomes, the data we consider in Section 5 involves a continuous response. Also, we have the issue of multiple testing.

### *3.1. Continuous response*

We observe the data $(\mathbf{X}_i, D_i)$, $i = 1, \ldots, m$, where $\mathbf{X}_i$ is the $n$-dimensional covariate vector and $D_i$ the binary phenotype for the $i$th individual. We formulate the following regression model relating $\mathbf{X}$ and $D$:

$$(3.1) \qquad E(X_{ki}|D_i, U) = \beta_{0k} + \beta_{1k}D_i + \gamma C_i,$$

where $(\beta_0, \beta_1)$ are regression coefficients, $C$ represents the confounder and $\gamma$ is the associated regression coefficient. Note that we assume no interaction between $C_i$ and $D_i$ in (3.1). Since $C$ is not observed, we fit the reduced model:

$$(3.2) \qquad E(X_{ki}|D_i) = \beta_{0k}^* + \beta_{1k}^* D_i.$$

We want to determine the relationship between $\beta_1$ and $\beta_1^*$. Note that integrating (3.1) with respect to the conditional distribution of $C$ given $D$ implies the following model:

$$(3.3) \qquad E[X_{ik}|D_i] = \beta_{0k} + \beta_{1k}D_i + \gamma E[C_i|D_i].$$

We now assume $C|D$ to be normal with mean $\mu_D$ and variance one. For this situation, plugging in the definition and equating regression coefficients for $D_i$ in (3.3) and (3.2) yields

$$\beta_{1k} = \beta_{1k}^* - \gamma(\mu_1 - \mu_0). \tag{3.4}$$

Thus, after the user specifies $\gamma$ and the difference $(\mu_1 - \mu_0)$, the estimate of $\beta_{1k}$, $\hat{\beta}_{1k}$, can be found quite easily by direct computation into (3.4) Since the inputs are constants, this has no impact on the standard errors of $\beta_{1k}$. One can then use $\hat{\beta}_{1k}/\hat{SE}(\hat{\beta}_{1k}^*)$ as the Wald statistic which when compared to a standard normal distribution yields a p-value. Again, the standard errors are unaffected by the sensitivity analysis parameters, and we can obtain p-values based on the Wald statistic.

### 3.2. Q-value/FDR adjustment for multiple testing

Based on the p-values calculated in the previous section, one can construct hypothesis-specific q-values (Storey and Tibshirani [31]). Q-values are defined to be the smallest FDR at which a test of hypothesis is significant, analogous to the p-value being the smallest level of significance at which a test of hypothesis is deemed to be significant. Here is the q-value construction algorithm:

1. Order the $G$ p-values as $p_{(1)} \leq p_{(2)} \leq \cdots \leq p_{(G)}$.
2. Construct a grid of $L$ $\lambda$ values, $\lambda_1, \ldots, \lambda_L$ and calculate

$$\hat{\pi}_0(\lambda_l) = \frac{\#\{p_j > \lambda_l\}}{G(1 - \lambda_l)},$$

$l = 1, \ldots, L$.
3. Fit a cubic smoothing spline to the values $\{\lambda_l, \hat{\pi}_0(\lambda_l)\}$, $l = 1, \ldots, L$.
4. Estimate $\pi_0$ by the interpolated value at $\lambda = 1$.
5. For the gene with the largest p-value the q-value is given by

$$q(p_{(G)}) = \min_{t \geq p_{(G)}} \frac{\hat{\pi}_0 G t}{\#\{p_j \leq t\}} = \hat{\pi}_0 p_{(G)},$$

and for $i = G - 1, G - 2, \ldots, 1$, $q(p_{(i)}) = \min(\hat{\pi}_0 G p_{(i)}/i, \hat{p}_{(i+1)})$. This guarantees that the q-values will be monotonically increasing as a function of p-values.

An alternative approach is to use the p-values in the Benjamini-Hochberg [2] procedure for controlling the false discovery rate:

(a) Let $p_{(1)} \leq p_{(2)} \leq \cdots \leq p_{(G)}$ denote the ordered, observed p-values.
(b) Find $\hat{k} = \max\{1 \leq k \leq G : p_{(k)} \leq \alpha k/G\}$.
(c) If $\hat{k}$ exists, then reject null hypotheses $p_{(1)} \leq \cdots \leq p_{(\hat{k})}$. Otherwise, reject nothing.

The major adjustment in these procedures is that we are accounting for potential adjustment of confounding, which impacts the p-values that get used in the multiple testing procedures.

Note that while we have used q-values to adjust for the multiple testing problem, the p-values can be input into any single-step algorithm for multiple testing. If we wished to control the familywise type I error level instead, we could use any of the procedures from Westfall and Young (1993) or some of the more recent methods from Van der Laan and Dudoit and colleagues (Van der Laan et al. [32, 33]).

One thing to note about the sensitivity analysis method described in Section 3.1 and is that it is effectively making an mean-shift adjustment to the theoretical null distribution. In particular, the standard errors of the estimators are not affected by the sensitivity analysis method of Lin et al. [19]. The multiple testing model that adjusts for confounding assumed here, in the notation of Section 2, is the following:

$$(3.5) \qquad T_i|\Theta \stackrel{iid}{\sim} \pi_0 f_0(t-\Theta) + (1-\pi_0)f_1(t),$$

where $\Theta$ denotes the sensitivity analysis inputs and are fixed constants specified by the user. One advantage of such a model for confounding is that it requires minimal user input and is easy to implement. A second advantage is that what are affected in this procedure is the estimate of $\pi_0$ and the actual q-values themselves. In settings where investigators are interested in the selection of hypotheses, these quantities might be all that is of interest.

In comparing this mixture model with (2.2) and (2.3), we find that the sensitivity analysis approach models confounding as a shift in the mean for the confounding null density relative to the theoretical null density. We allow for nonconstant variance across the genome, in contrast to the approach of Devlin and Roeder [10]. In addition, unlike Devlin and Roeder [10], we model the bias due to confounding. By comparison, the Pritchard and Rosenberg [23] and Efron [11] approaches would allow for both a mean and variance adjustment for the null hypothesis density. It appears that combining the approach here with that of genomic control should provide an answer that is closer to the Pritchard and Rosenberg [23] proposal, if supplemental data were available.

### *3.3. Shrinkage estimation for the q-value*

We now heuristically describe an argument, given in more detail in Ghosh [16], that shows how the q-value can be motivated as a shrinkage estimator in the multiple testing problem. Consider model (2.1) again, where the test statistics are the p-values. Then one could construct shrinkage estimators of the p-values where they are shrunk towards the components of the mixture model defined by (2.1). If we assume that the density of the null hypothesis is a point mass at one and that of the alternative is zero, then it is shown in Ghosh [16] that a shrinkage estimator of the p-value in this problem is lower bounded by the q-value. Thus, the q-value approach advocated by Storey [28] enjoys a nice shrinkage property. However, the definition of the q-value given by Storey [28] is appropriate to the situation of 0/1 loss (misclassification error). The shrinkage estimation procedure by Ghosh [16] could generalize easily to other loss functions, which might be more appropriate in other scientific contexts.

### **4. Estimators of association in the presence of confounding**

A limitation of all these approaches is that they do not produce multiple-testing adjusted estimates of association or confidence intervals. We now discuss the problem of how to estimate association measures in the multiple testing problem in the presence of confounding. While the multiple testing problem has been considered quite intensively in the statistical literature, that of estimating associations has not.

This was pointed out recently by Prentice and Qi [21] in a genomewide study example. Our approach is based on the idea that we can use (2.1) to define estimation targets under the null and alternative hypotheses. One then can define Empirical Bayes estimators that shrink the observed test statistic towards each target with appropriate mixing weights.

For estimation, James-Stein (James and Stein [18]) estimators are used. We now construct the double shrinkage estimators using the test statistic model (2.1) where $F = \pi_0 F_0 + (1-\pi_0) F_1$. To do this, we utilize the density estimation method proposed by Efron [11]. Note from (2.1) that we have

$$\pi_1 f_1(t) = f(t) - \pi_0 f_0(t).$$

We can estimate $f(t)$ by applying density estimation methods to $T_1, \ldots, T_G$. For estimation of $\pi_0 f_0(t)$, the zero assumption in Efron [11] is utilized. What this means is that most test statistics with a value near zero comes from the null distribution component. The assumption, combined with a normal-based moments matching technique and density estimation procedures as described in Efron [11], allows one to obtain an estimate of $\pi_0$ and $f_0(t)$. Given them, we then obtain an estimate of $\pi_1$ and $f_1(t)$ by simple subtraction. Based on the estimates of $f_0(t)$, $f_1(t)$ and $\pi_0$, we can estimate $T_1^{JS}, \ldots, T_n^{JS}$ by

(4.1) $$\hat{T}_i^{JS} = \hat{\pi}_0(T_i) \hat{T}_{0i}^{JS} + \{1 - \hat{\pi}_0(T_i)\} \hat{T}_{0i}^{JS},$$

where

$$\hat{T}_{0i}^{JS} = T_i - \left[1 \wedge \frac{G-2}{\sum_{i=1}^G (T_i - \hat{\mu}_0)^2}\right] (T_i - \hat{\mu}_0),$$

$$\hat{T}_{1i}^{JS} = T_i - \left[1 \wedge \frac{G-2}{\sum_{i=1}^G (T_i - \hat{\mu}_1)^2}\right] (T_i - \hat{\mu}_1),$$

$$\hat{\pi}_k(t) = \frac{\hat{\pi}_k \hat{f}_k(t)}{\hat{\pi}_0 \hat{f}_0(t) + (1-\hat{\pi}_0) \hat{f}_1(t)},$$

$\hat{\mu}_0 = \int t d\hat{F}_0(t)$ and $\hat{\mu}_1 = \int t d\hat{F}_1(t)$.

Before discussing the issue of confidence intervals, note that we are using the model framework of Efron [11] to deal with confounding, i.e., model (2.4). As mentioned before, estimation in this model requires no supplemental data on known null hypotheses.

Along with estimators of the parameters that adjust for multiple testing, it is useful to have confidence intervals that also account for the multiple testing phenomenon. To calculate the confidence intervals, we will use the following simulation-based algorithm:

1. Estimate $\pi_0$, $f_0$ and $f_1$ using the algorithm above.
2. Construct the double shrinkage estimators of $\mu_1, \ldots, \mu_G$, $\hat{T}_1^{JS}, \ldots, \hat{T}_G^{JS}$.
3. Sample with replacement $G$ observations $\mu_1^*, \ldots, \mu_G^*$ from the estimated density $\hat{\pi}_0 \hat{f}_0(t) + (1-\hat{\pi}_0) \hat{f}_1(t)$ and generate $T_{1,1}^*, \ldots, T_{G,1}^*$, where $T_{i,1}^* \sim N(\hat{T}_i^{JS}, 1 + \hat{\sigma}_*^2)$, where $\hat{\sigma}_*^2$ is the empirical variance of the $\mu^*$'s sampled in Step 2.
4. Repeat Step 3 $B$ times. Use the empirical distribution of $T_{i,1}^*, \ldots, T_{i,B}^*$ to calculate confidence intervals for $\mu_i$, $i = 1, \ldots, G$.
5. Multiply by the gene-specific standard deviations to get confidence intervals on the scale of the parameter.

Based on the bootstrap distributions, we can construct confidence intervals for each of the components of $\mu \equiv (\mu_1, \ldots, \mu_G)$. We have chosen to use equal tail confidence intervals, e.g. a 95% confidence interval would be based on the 2.5th and 97.5th percentiles of the empirical distribution of $T_i^*$.

## 5. Numerical example: Microarray profiling study in prostate cancer

We now illustrate the proposed methodologies using two genomic studies. The first comes from prostate cancer gene expression profiling experiment reported in Varambally et al. [35]. The investigators used cDNA microarrays containing 9984 genes to profile tissue samples from various stages of prostate cancer (normal adjacent prostate, benign prostatic hyperplasia, localized prostate cancer, advanced metastatic prostate cancer).

Measurements were made on 9984 genes for 101 individuals, of whom 79 have cancers and 22 do not. Before analyzing the data, we filtered out genes that had a sample variation less than 0.05 across all samples. This left a total of $G \equiv 6253$ genes available for analysis.

First, a q-value analysis (Storey [28]) was done using the unpooled two-sample t-test; this is reported in Figure 1. For this analysis, we assumed a theoretical null distribution of $N(0,1)$ for each t-statistic. One point of note from the plot is the estimate of $\pi_0$, the fraction of null hypotheses estimated to be true, is 0.49. As a result, we are finding a large proportion of genes to be differentially expressed.

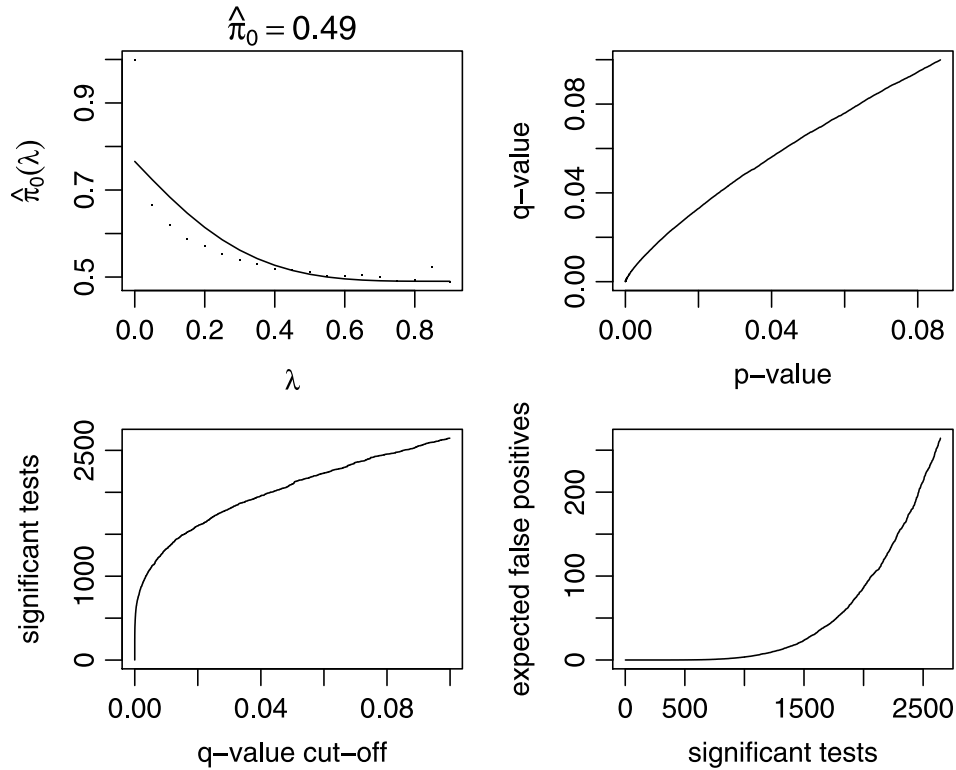

FIG 1. *Results of the q-value analysis (Storey [28]) for prostate cancer gene expression data.*

TABLE 2
*Results of sensitivity analyses for gene expression data*

| $\gamma$ | $(\mu_1 - \mu_0)$ | $\hat{\pi}_0$ | Number q-values $\leq 0.05$ |
|---|---|---|---|
| 0 | 0 | 0.49 | 2099 |
| 1.5 | 0.01 | 0.86 | 1762 |
|  | 0.1 | 0.47 | 613 |
|  | 0.3 | 0.05 | 6227 |
|  | 0.5 | 0.01 | 6253 |
|  | 1 | 0.006 | 6253 |
| 0.1 | 0.01 | 0.49 | 2415 |
|  | 0.1 | 0.78 | 1902 |
|  | 0.3 | 0.86 | 1466 |
|  | 0.5 | 0.80 | 1154 |
|  | 1 | 0.69 | 586 |

As argued by Ghosh and Chinnaiyan [17], there is a variety of confounders that might cause the apparent observed difference between the cancer and noncancer samples. The implication is that the standard normal distribution might not be the correct reference null distribution to use. We tried adjustments to the t-statistic based on the sensitivity analysis approach. First, we assumed a confounder that was continuous and tried various values of $(\gamma, \mu_1, \mu_0)$ to obtain estimates of $\pi_0$ and the q-values. Note that what matters is the difference in means $\mu_1 - \mu_0$, so we considered changes of $0.01, 0.1, 0.3, 0.5$ and 1. We have aggregated the sensitivity results into Table 2. What we find is that the estimate of $\pi_0$ is extremely sensitive to confounders. Interestingly, as the product of $\gamma$ and $(\mu_1 - \mu_0)$ decreases, we find that the estimate of the proportion of true null hypotheses increases, which leads to a decrease in the number of genes being called significant if we use a cutoff of 0.05, say. However, we also see that there is no simple relationship between the amount of confounding and the estimate of $\pi_0$. The estimate of number of genes being called significant is also sensitive to the estimate of $\pi_0$. For example, there is very little difference in the estimates of $\pi_0$ for the unadjusted and $(\gamma, \mu_1 - \mu_0) = (0.1, 0.01)$ situation, yet we are calling approximately four hundred more genes significant in the second scenario. This emphasizes the need to have good estimates of $\pi_0$.

Next, we utilized the Efron [11] method for estimating the local false discovery rates; the results are presented in Figure 2. Again, this procedure corresponds to a data-driven adjustment to null hypotheses for confounding. Now note that the estimate of $\pi_0$ from the Efron [11] procedure is 0.88, almost 80% larger than the estimate using the method of Storey [28]. This fits into our intuition that adjusting for confounding should decrease the number of significant genes. We next calculated the double-shrinkage estimators for the t-statistics using the method in Section 3.1. We then combined an FDR-thresholding procedure with the inference procedures described in the paper. This was done in the following manner. We applied the q-value procedure of Storey [28] and selected the genes with the twenty smallest q-values. We reported the effects and associated 95% confidence intervals in Table 3. Note that the 95% CIs should be conservative in that we have ignored the selection process (i.e. taken the twenty genes with the smallest q-values).

## 6. Discussion

In this article, we have addressed the issue of multiple testing procedures in the presence of confounding. First, we sought to unify proposals from the statistical and genetics literature on the problem. Doing so clarifies and provides a stronger

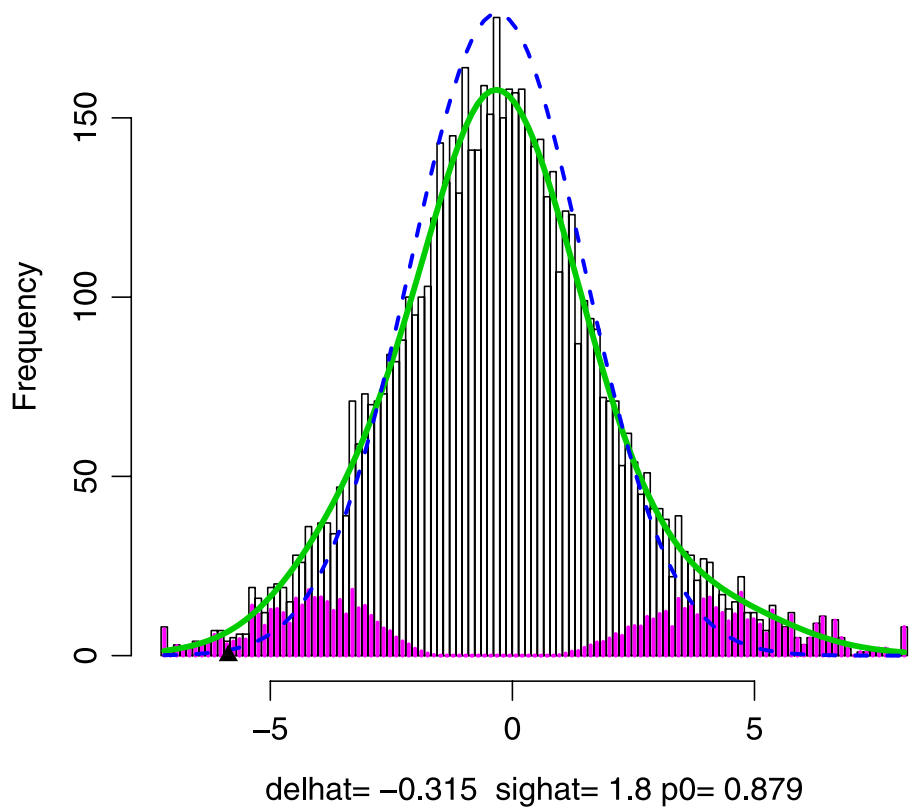

delhat= –0.315  sighat= 1.8 p0= 0.879

FIG 2. *Results of the t-test statistics from the prostate cancer gene expression data using the local false discovery rate estimation procedure of Efron [11].*

TABLE 3
*Results for top 20 genes for prostate cancer gene expression data*

| Gene name | Estimator S | $\hat{\beta}$ | 95% CI |
|---|---|---|---|
| Transforming growth factor, beta 1 | 6.99 | 0.30 | (0.09, 0.52) |
| Fer-1-like 3, myoferlin (C. elegans) | 8.36 | 0.55 | (0.14, 0.93) |
| O-linked N-acetylglucosamine transferase | −9.11 | −0.51 | (−0.88, −0.18) |
| Human calmodulin-I (CALM1) mRNA | 6.87 | 0.56 | (0.17, 0.98) |
| Hepsin (transmembrane protease, serine 1) | −6.89 | −0.95 | (−1.62, −0.30) |
| Caveolin 2 | 7.44 | 0.53 | (0.16, 0.85) |
| Ras homolog gene family, member B | 6.7 | 1.41 | (0.42, 2.40) |
| Zinc finger protein 36, C3H type-like 1 | 6.76 | 0.52 | (0.16, 0.87) |
| Hepatic leukemia factor | 6.72 | 0.46 | (0.11, 0.78) |
| Phosphatidic acid phosphatase type 2B | 6.59 | 0.34 | (0.08, 0.60) |
| Multiple endocrine neoplasia I | −6.51 | −0.31 | (−0.54, −0.08) |
| Endothelin receptor type A | 6.47 | 0.51 | (0.12, 0.92) |
| Transforming growth factor, beta receptor III | 6.38 | 0.54 | (0.10, 0.98) |
| Growth factor receptor-bound protein 2 | −6.31 | −0.42 | (−0.77, −0.12) |
| Hermansky-Pudlak syndrome 1 | 6.32 | 0.34 | (0.07, 0.58) |
| Dickkopf homolog 3 (Xenopus laevis) | 6.29 | 0.34 | (0.09, 0.61) |
| Phosphatidic acid phosphatase type 2B | 6.24 | 0.32 | (0.06, 0.57) |
| Tissue inhibitor of metalloproteinase 3 | 6.2 | 0.55 | (0.12, 1.00) |
| Hypothetical protein hCLA-iso | −6.13 | −0.38 | (−0.70, −0.09) |
| Zinc finger protein 36, C3H type-like 2 | 6.08 | 0.34 | (0.08, 0.62) |

justification for preferring the empirical null distribution that Efron [11] has recently advocated. It is also clear from the discussion that what the methods of Devlin and Roeder [10] and Pritchard and Rosenberg [23] provide are adjustments to the theoretical null, using supplemental data and under various assumptions on the nature of confounding and its effects on location and scale parameters.

Second, we have proposed two analytical approaches. The first is a sensitivity analysis approach using a methodology described by Lin et al. [19], which is then plugged into standard FDR methodology. As shown in the article, this can be viewed as a mean-shift adjustment to the theoretical null. The second procedure in the article we pursue is one using Empirical Bayes methodology. This involves utilizing the mixture model for hypothesis testing in the presence of confounding proposed by Efron [11] and constructing James-Stein estimators of the test statistic that shrinkage towards each target (that specified under the null and that under the alternative) with data-dependent weights. What is novel here is that the mean value of the distribution of the test statistic under the null is also estimated as part of the procedure. This leads to calibration of the test statistics under the empirical null hypothesis (Efron [11]) rather than the theoretical null hypothesis. We also develop confidence intervals in the multiple testing setting, which has not been discussed very much in the literature, with the major exception being the work on FDR-controlling confidence intervals by Benjamini and Yekutieli [5].

The methodology proposed here is fairly general. The sensitivity analysis regression models considered here are a subset of those considered by Lin et al. [19]; the methodology proposed here could apply more generally to censored outcomes through proportional hazards regression models and count responses using Poisson regression models. The double shrinkage estimation methodology in Section 4 requires having Wald-type estimators.

While there have been Empirical Bayes methods for multiple testing proposed in the literature (Efron et al. [12], Datta and Datta [9] and Ghosh [16]), most of this work has focused on hypothesis testing and selection of hypotheses. While this is useful in screening problems, it might also be of interest to report estimated effects and confidence intervals corresponding to the rejected null hypotheses. A major potential advantage of Empirical Bayes (and more generally, fully Bayesian) methods in this setting is that parameter estimates and confidence intervals will be unaffected by the selection of which hypotheses to reject, provided the testing procedure is Bayesian. This is an area that is currently under investigation.

**Acknowledgments.** The author would like to thank Tom Nichols and Trivellore Raghunathan for helpful discussions.